\documentclass{aa}  

\usepackage{graphicx}
\usepackage{txfonts}
\usepackage{lipsum}
\usepackage{subcaption}         % necessary for continued figures, example in section 3
\usepackage{lscape}             % to rotate a single page table, example in appendix.
\usepackage{placeins}           % useful with \FloatBarrier, to keep 
\usepackage[colorlinks=true, citecolor=teal]{hyperref}
\usepackage{xcolor}

\begin{document}

%%%%%%%%%%%%%%%%%%%%%%%%%%%%%%%%%%%%%%%%
% if you use custom commands in your title,
% ensure to check your title when submitting!
%%%%%%%%%%%%%%%%%%%%%%%%%%%%%%%%%%%%%%%%
   \title{Constraining the model-based uncertainties of asteroseismic magnetic field measurements in red giants}
   % \subtitle{Subtitle}

%%%%%%%%%%%%%%%%%%%%%%%%%%%%%%%%%%%%%%%%
% Please separate each author with the \and command
%
% Please do not include ORCIDs next to author names.
% Only ORCIDs authenticated by individual authors in EDPS
% editorial system will be taken into account.
% ORCIDs included here will be removed.
%%%%%%%%%%%%%%%%%%%%%%%%%%%%%%%%%%%%%%%%

   \author{L. Buchele \inst{1}
      \and L. Bugnet \inst{1}
      \and N. Muntean \inst{1}
      \and E. Hatt \inst{1,2}% Others? 
        }

   \institute{Institute of Science and Technology, 3400, Klosterneuburg, Austria
   \and School of Physics \& Astronomy, University of Birmingham, Edgbaston, Birmingham B15 2TT, UK}

   % \date{Received September 30, 20XX}

% \abstract{}{}{}{}{}
% 5 {} token are mandatory
 
  \abstract % current draft is 300 words long -- exactly at the limit. 
   {Magnetic fields in the radiative interiors of red giants are measured using shifts in stellar oscillation frequencies. However, in the asymptotic framework, converting an observed frequency shift into a radial magnetic field strength requires knowing the global magnetic sensitivity. The parameter which describes this sensitivity (also called the core structure parameter) must be inferred from stellar models, which introduces a potential source of uncertainty.}
   {This work seeks to understand how the global magnetic sensitivity depends on stellar properties such as mass and metallicity, and to quantify the model-based uncertainty on magnetic field measurements. We also explore which stellar properties are key to finding a precise and accurate estimate of the global magnetic sensitivity parameter.} 
   {Using MESA models, we examine how the global magnetic sensitivity changes with mass, metallicity, and age. We then create a sample of synthetic stars and test how well our grid-based fitting method recovers the sensitivity parameter. We consider different grid construction approaches and the choice of which observational properties are used in the fitting process.}
   {We find that the global magnetic sensitivity shows a stronger dependence on mass for higher mass models and a stronger metallicity dependence for lower metallicity models. Our fitting methods recover the underlying sensitivity parameter well, with a model-based uncertainty of 10\% when precise metallicity measurements are used. We apply our procedure to stars with existing magnetic field measurements. In most cases, the dominant source of uncertainty remains observational, although such precise modeling can significantly reduce the magnetic field uncertainty for stars with exceptional data.}
   {With careful fitting, stellar models yield accurate values for the global magnetic sensitivity. We recommend that future work obtain the global magnetic sensitivity  using both asteroseismic and high-quality spectroscopic constraints. Under these conditions, we recommend adopting a model-based uncertainty of 10\% on the sensitivity parameter.}


   \maketitle
   \nolinenumbers

%%%%%%%%%%%%%%%%%%%%%%%%%%%%%%%%%%%%%%%%%%%%%%%%%%%%%%%%%%%%%%
\section{Introduction}
High-quality space-based asteroseismic data provide avenues to explore stellar physics across many different kinds of stars. In red giants, the primary focus of this work, it has been known for some time that classic stellar evolution models \citep[e.g.][]{cantiello_angular_2014, ceillier_surface_2017, eggenberger_rotation_2022} do not reproduce the measured core and envelope rotational rates \citep{beck_fast_2012, mosser_spin_2012, deheuvels_seismic_2014, di_mauro_inernal_2016, triana_internal_2017, gehan_core_2018, li_asteroseismic_2024}. One of the proposed solutions to this problem is that magnetic fields transport angular momentum \citep{cantiello_angular_2014, spada_angular_2016, eggenberger_asteroseismology_2019, fuller_slowing_2019, gouhier_angular_2022, eggenberger_rotation_2022, moyano_asteroseismology_2023, meduir_angular_2024}. Of course, the efficiency of angular momentum transport provided by a magnetic field depends on the properties of the field \citep[e.g.][]{takahashi_modeling_2021}, and so ensuring that measured field strengths are accurate is a key part of addressing the angular momentum problem. Thanks to the high-quality asteroseismic data provided by the \textsl{Kepler} mission, it is now possible to measure the average radial magnetic field strengths in red giant radiative interiors \citep{li_magnetic_2022, deheuvels_strong_2023, li_internal_2023, hatt_asteroseismic_2024, villate_seismic_2026}, using the perturbative theory for weak to moderate magnetic fields \citep{bugnet_magnetic_2021, mathis_probing_2021, li_magnetic_2022,  bugnet_magnetic_2022, mathis_asymmetries_2023, bhattacharya_detectability_2024, das_unveiling_2024}. In addition, the effect of stronger near-critical fields \citep{fuller_asteroseismology_2015, loi_effect_2020} can be studied using the traditional approximation of rotation and magnetism \citep[TARM,][]{Dhouib2022, rui_asteroseismic_2024, ligieres_pertubative_2024, deheuvels_near-critical_2026}. In our study, we remain within the perturbative framework and consider the magnetic field strength small enough not to be affected by near-critical effects \citep[see][for validity criteria]{bugnet_magnetic_2021}.

Both rotational profiles \citep[e.g.][]{beck_fast_2012} and such internal magnetic fields \citep[e.g.][]{bugnet_magnetic_2021, li_magnetic_2022} can be measured by observing shifts in the oscillation mode frequencies. The average core rotation can be determined directly from these shifts \citep[e.g.][]{beck_fast_2012, mosser_spin_2012, gehan_core_2018, li_asteroseismic_2024} or rotational inversions can be done using a stellar reference model to obtain both core and envelope rotation rates \citep[e.g][]{deheuvels_seismic_2014, di_mauro_inernal_2016, ahlborn_improved_2022}. In the case of magnetic fields, stellar models are also necessary to convert measured magnetic splitting into average magnetic field strengths \citep{mathis_probing_2021, li_magnetic_2022}. While some work has been done to understand the uncertainty introduced by stellar modeling into rotational inversions \citep{ahlborn_robustness_2025}, less attention has been paid to the potential model dependence of magnetic field measurements. \cite{bugnet_magnetic_2021} investigated the dependency of the magnetic signature on mass and metallicity before any observational magnetic signature had been detected, and showed that the stellar properties slightly affect the sensitivity of the modes to the magnetic fields. Now that a few dozen magnetic red giants are being characterized through the use of stellar models, it becomes important to estimate the modeling bias in the context of observational measurements.

The main method of measuring weak to moderate red giant internal magnetic fields relies on measuring the g-mode frequency perturbation, $\delta \nu_{\rm{mag}}$, attributable to the magnetic field \citep[][]{bugnet_magnetic_2021}. In practice, it is common to take $\delta \nu_{\rm{mag}}$ as the magnetic shift for pure $g$ modes at the frequency of maximum power, $\nu_{\rm{max}}$. The value of $\delta \nu_{\rm{mag}}$ is obtained from the observed mixed dipole modes by fitting  \citep{li_magnetic_2022}: 
\begin{equation}
    \label{equ:m0_shifts} 
    \delta \nu_{\rm{mag}, m=0} = \zeta \: (1-a) \delta \nu_{\rm{mag}} \left(\frac{\nu_{\rm{max}}}{\nu}\right)^3,     
\end{equation}
\begin{equation}
    \label{equ:m1_shifts} 
    \delta \nu_{\rm{mag}, m=\pm 1} = \zeta \: \left(1+ \frac{a}{2}\right) \delta \nu_{\rm{mag}} \left(\frac{\nu_{\rm{max}}}{\nu}\right)^3,     
\end{equation}
where $m$ is the azimuthal order, $\zeta$ is the gravity mode inertia, $\nu$ is the mode frequency, and $a$ is a parameter that depends on the geometry of the field. In order to constrain both $\delta \nu_{\rm{mag}}$ and $a$, all three components ($m=-1,0,1$) of the dipole modes must be visible. In Equations~\ref{equ:m0_shifts}~and~\ref{equ:m1_shifts}, $\zeta$ accounts for the fact that only the $g$ part of the mixed mode is shifted by the magnetic field. Alternatively, the mixed mode can be decomposed into the underlying $p$ and $g$ components, and the magnetic shift applied directly to the $g$ component which is then re-coupled to the pure $p$ mode, as was done by \citet{hatt_asteroseismic_2024}. Once an overall value of $\delta \nu_{\rm{mag}}$ has been obtained, this shift is then attributed to the dominant term of the magnetic field effect, which is the radial component of the field \citep{bugnet_magnetic_2021, mathis_probing_2021, li_magnetic_2022}. \cite{li_magnetic_2022} further expressed the asymptotic link between the average squared radial field strength and the magnetic shift as:
\begin{equation} 
    \label{equ:shift_to_strength} 
    \left<B_r^2\right>
= \frac{16 \pi^4 \mu_0}{\mathcal{I}} \nu_{\rm{max}}^3 \delta \nu_{\text{mag}}, 
\end{equation}
where $\mu_0$ is the vacuum permeability and $\mathcal{I}$ parametrizes the global magnetic sensitivity. In previous works \citep[e.g.][]{li_magnetic_2022}, $\mathcal{I}$ was called the core structure parameter, however we refer to it as the global magnetic sensitivity to make its usage clearer. If the magnetic shift is calculated at an arbitrary frequency rather than $\nu_{\rm{max}}$, then this frequency replaces $\nu_{\rm{max}}$ in Equation~\ref{equ:shift_to_strength}. $\mathcal{I}$ is what allows the shift to be converted into a measured average field strength, and it is defined as
\begin{equation}
    \label{equ:I_def} 
    \mathcal{I} = \frac{\displaystyle{\int_{r_i}^{r_o} \left(\frac{N}{r}\right)^3 \frac{{\rm{d}}r}{\rho}}}{\displaystyle{\int_{r_i}^{r_o} \left(\frac{N}{r}\right) {\rm{d}}r}}, 
\end{equation}
where $N$ is the buoyancy frequency (also called Brunt–Väisälä frequency), $\rho$ is the density, and the integration bounds $r_i$ and $r_o$ are the inner and outer turning points of the g-mode cavity, respectively. 

$\mathcal{I}$ cannot be directly observed and is therefore the only non-observational component leading to the estimate of the magnetic field strength in the perturbative framework \citep{li_magnetic_2022, li_internal_2023, deheuvels_strong_2023, hatt_asteroseismic_2024, villate_seismic_2026}. Instead, it must be obtained from a stellar model, typically the best-fit model to other observed stellar properties. Thus, there is some degree of model-dependence in the final estimate of the internal magnetic field strength from this simplified formulation. This dependence comes from the choices made both when constructing the models and when finding the best-fit model. 

Several works have used the method outlined above to measure internal magnetic fields in red giants, and each work has taken a different approach to quantifying the uncertainty introduced by inferring $\mathcal{I}$ from stellar models. \citet{li_magnetic_2022} used the $\mathcal{I}$ values of the models from their grid (which varied mass, metallicity, and age) that reproduced both the large frequency separation ($\Delta \nu$) and the asymptotic period spacing of the dipole modes ($\Delta \Pi_1$). This is also the method used by \citet{deheuvels_strong_2023}. \citet{li_internal_2023} also took the $\mathcal{I}$ value of the best fitting model, although in this case the model was fit to the stellar effective temperature ($T_{\rm{eff}}$), luminosity ($L$), metallicity ([Fe/H]), radial-mode frequencies, and $\Delta \Pi_1$. For all but one star studied by \citet{villate_seismic_2026}, $\mathcal{I}$ was obtained from the model that best fit the observed $\nu_{\rm{max}}, \Delta \nu,$ and $\Delta \Pi_1$ values. The one exception was a star whose $\Delta \Pi_1$ value was below the $\Delta \nu$-$\Delta \Pi_1$ sequence expected from single star evolution. In this case \citet{villate_seismic_2026} found a best-fit model using only the observed $\Delta \Pi_1$. In all four works, the value of $\mathcal{I}$ has been taken from a best-fit model without including any model-based uncertainty. 

In contrast, \citet{hatt_asteroseismic_2024} do include uncertainty on $\mathcal{I}$ in the field strengths they report. Their $\mathcal{I}$ values are found by fitting the observed $T_{\rm{eff}}$, [Fe/H], surface gravity, mass \citep[M, obtained from scaling relations in][]{yu_asteroseismology_2018}, $\Delta \Pi_1$, $\nu_{\rm{max}}$, $\Delta \nu$, and radial model frequencies. They find that the distribution of $\mathcal{I}$ across all of their best-fit models is approximated by a Gaussian with a width of 30\% of the mean, and so they adopt a 30\% uncertainty on $\mathcal{I}$.

These two approaches, assuming either no uncertainty on $\mathcal{I}$ or an uncertainty defined by the spread across all stars, represent the extremes of how model-based uncertainty can be accounted for. This work aims to find a middle ground, which acknowledges that while our stellar models likely do not fully represent the stars we observe, we do have some constraints on the properties of individual stars. To find this middle ground, we seek to understand how the value of $\mathcal{I}$ depends on the choices made when constructing stellar models and the procedure used to find the best-fit model. For this, we use two grids of models, described in Sect.~\ref{sect:models}. We focus on one grid of models to explore how $\mathcal{I}$ depends on the stellar parameters in Sect.~\ref{sect:params}. This provides us with some insight into what should be considered when finding an appropriate value of $\mathcal{I}$. In Sect.~\ref{sect:best-fit}, we test how well known values of $\mathcal{I}$ are recovered using our each grid of models and various sets of observational parameters. We then use our models and fitting procedure to infer $\mathcal{I}$ for red-giant branch stars with previously measured average magnetic field strengths in Sect.~\ref{sect:compare}. Finally, we present our conclusions and recommendations in Sect.~\ref{sect:conc}.

%%%%%%%%%%%%%%%%%%%%%%%%%%%%%%%%%%%%%%%%%%%%%%%%%%%%%%%%%%%%%%
\section{Models} \label{sect:models} 
This work uses two different grids of stellar models generated using version \texttt{r24.08.1} of the MESA stellar evolution code \citep{paxton_modules_2011, paxton_modules_2013, paxton_modules_2015, paxton_modules_2018, paxton_modules_2019, jermyn_modules_2023}. The parameters that were kept constant between both grids are detailed in Appendix~\ref{app:MESA}. In general, our modeling choices follow those of \citet{li__beyond_2025}, with the most significant change being that we use the exponential form of overshoot \citep{herwig_evolution_2000}. Both grids cover the mass range $0.8 \: {\rm{M}}_\odot  \leq M  \leq 1.8 \: {\rm{M}}_\odot$ and initial metallicity range $-1 \leq {\rm{[Fe/H]}}_{\rm{init}} \leq 0.5$. In our grid labeled Linear, we adopted a linear sampling approach where the models are evenly spaced in mass and initial metallicity with step sizes of $0.05 \: {\rm{M}}_\odot$ and 0.25 dex, respectively, leading to 147 tracks. In addition, the helium enrichment law was fixed such that the initial helium mass fraction was Y$_{\rm{init}} = 0.24 + 2$Z$_{\rm{init}}$, where Z$_{\rm{init}}$ is the initial metal mass fraction. We also adopted a fixed mixing length parameter ($\alpha_{\rm{mlt}}$ =2) and fixed overshoot parameter ($f_{ov} = 0.015$). For our grid labeled Sobol, we adopted a quasi-random sampling of M, Y$_{\rm{init}}$, [Fe/H]$_{\rm{init}}$, $\alpha_{\rm{mlt}}$, and $f_{ov}$ using a Sobol sequence (\citealt{sobol_1967}, see also Appendix B of \citealt{bellinger_fundamental_2016}). We sample mass and metallicity in the same ranges as our linear grid and use the following ranges for the additional free parameters: Y${\rm{init}} \in [0.24, 0.33], \, \alpha_{\rm{mlt}} \in [1.3, 2.6], f_{ov} \in [0.004, 0.06]$. As this grid has three additional free parameters, we also increase the number of tracks to 2048 to ensure sufficient sampling of the space.

We choose to use two grids for several reasons. First, although Sobol sequences provide better coverage of multidimensional parameter spaces, it is harder to isolate the effects of changing a single parameter because the quasi-random sampling returns stellar parameters that are always slightly different. This makes it more difficult to, for example, examine the difference between tracks of varying masses but fixed metallicity. This is why we use our linear grid to explore how $\mathcal{I}$ depends on the stellar parameters in Sect.~\ref{sect:params}. The second reason we use these two grids is to better represent the modeling procedures used in previous works. Although there are some differences in the physical prescriptions we use, the mass and metallicity values we choose in our Linear grid are the same as the grid used by \citet{hatt_asteroseismic_2024}. Our Sobol grid was instead constructed to be closer to the grid used by \citet{li_internal_2023}, which was initially described in \citet{li_t_asteroseismology_2022}.

\begin{figure*} % Figure placed here so that it appears on page 3
    \centering
    \includegraphics{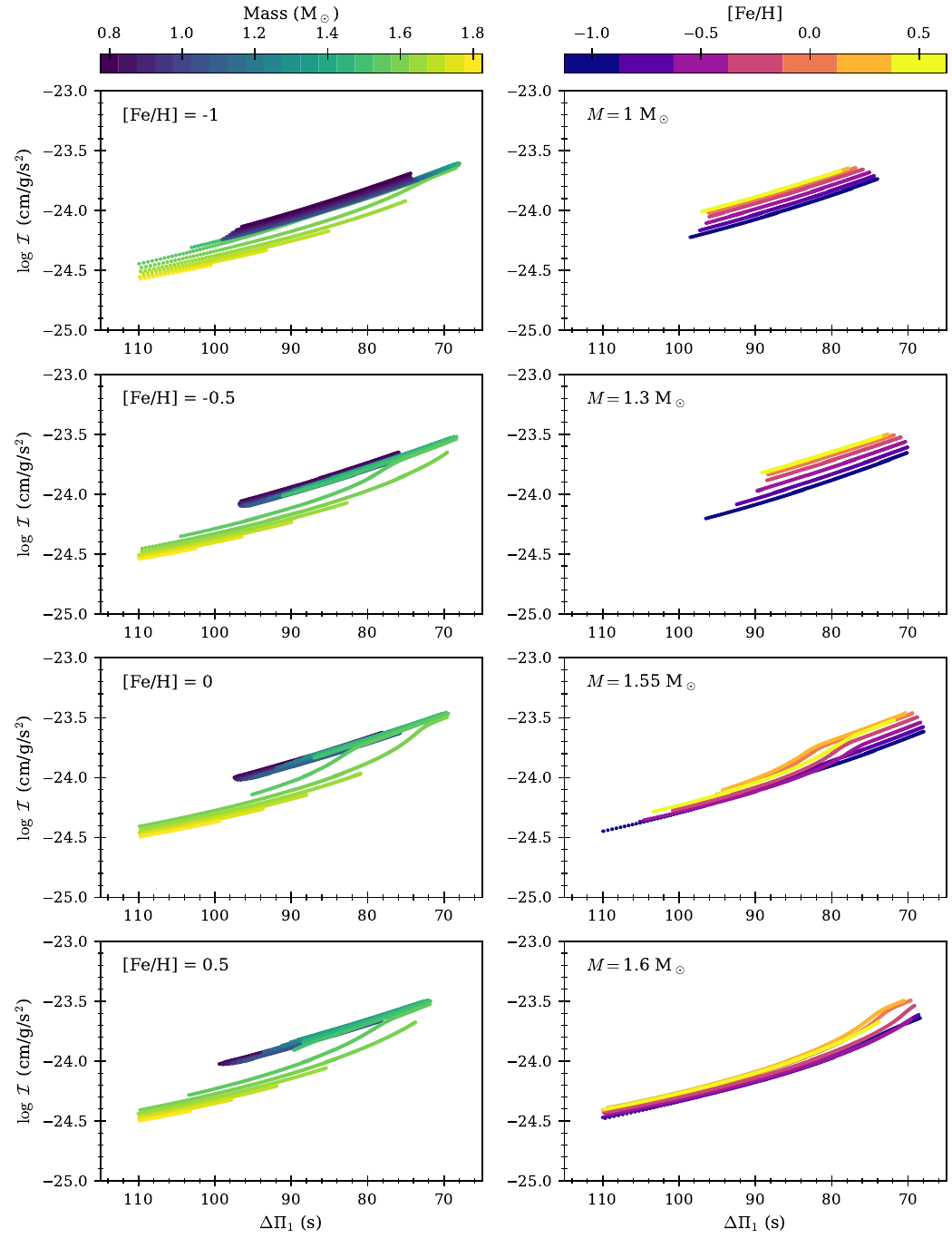}
    \caption{Dependence of the global magnetic sensitivity, $\mathcal{I}$  (also called the core structure parameter), on stellar parameters for models in our linear grid. All panels show $\mathcal{I}$ as a function of $\Delta \Pi_1$ for a subset of tracks in our linear grid. In the left column, the tracks shown in each panel have the same metallicity but vary in mass. In the right column, each panel shows tracks of the same mass but different metallicity.  In all panels, the $\Delta \Pi_1$ axis is reversed, such that a given evolutionary track evolves from left to right.}
    \label{fig:mid_os_M_FeH}
\end{figure*}

Along each track, we calculated $\Delta \nu$ using the radial mode frequencies obtained from the stellar oscillation code GYRE \cite{townsend_gyre_2013}, $\Delta \Pi_1$ using the asymptotic form
\begin{equation} 
\label{equ:DPi} 
\Delta \Pi_1 = \sqrt{2}\, \pi^2 \left(\int_{r_i}^{r_o} \frac{N}r \:  {\rm{d}} r \right)^{-1}, 
\end{equation}
and $\mathcal{I}$ according to Equation~\ref{equ:I_def}. We provide more details about these calculations in Appendix~\ref{app:MESA}, including an important change from the MESA default calculation of $\Delta \Pi_1$. From each grid, we selected models with $100 \: \mu{\rm{Hz}} \leq \nu_{\rm{max}} \leq 277 \: \mu{\rm{Hz}}$ and~$\Delta\Pi_1~<~110$~s, which corresponds roughly to the range of parameters in the sample of  \citet{hatt_asteroseismic_2024}.

\begin{figure*}
    \centering
    \includegraphics[width=\linewidth]{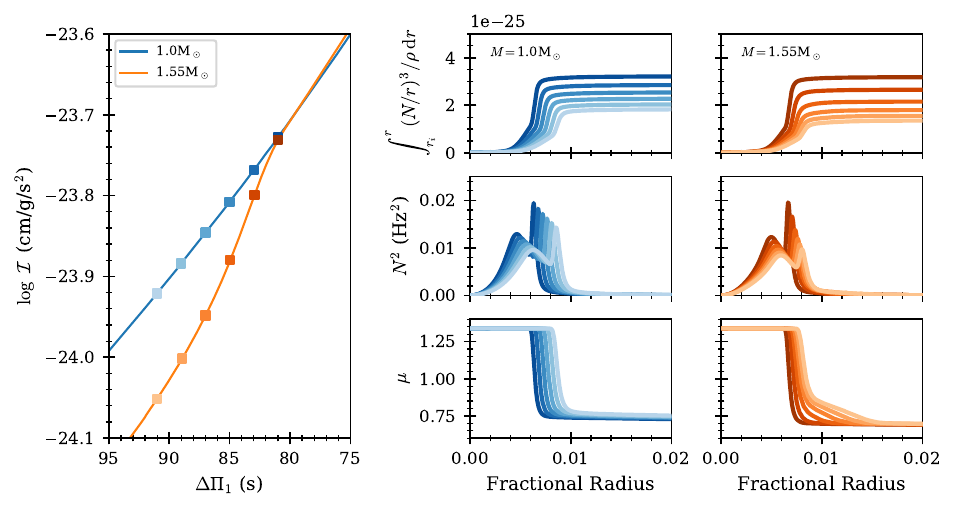}
    \caption{Connection between mean molecular weight profiles and $\mathcal{I}$ for models with solar metallicity of two different masses. The left-most panel shows the evolution of $\mathcal{I}$ for the two tracks, with squares indicating the selected models plotted in the second and third columns. The top panels in these columns show the cumulative integral of the numerator in Equation~\ref{equ:I_def} for each selected model of each track. The middle and bottom panels in these two columns show the $N^2$ and $\mu$ profiles, respectively. In the second and third columns, the darker lines correspond to more evolved models. We note that although the burning shell is moving outwards in mass, the contraction of the core results in the burning shell moving inward in radius.}
    \label{fig:N2_mass}
\end{figure*}

\section{Dependence of $\mathcal{I}$ on stellar parameters}  \label{sect:params} 
In order to explore how modeling choices may affect an inferred value of $\mathcal{I}$, it is first useful to understand how $\mathcal{I}$ depends on the stellar properties. For this, we use only the models in our linear grid. In Fig.~\ref{fig:mid_os_M_FeH}, we show how the evolution of $\mathcal{I}$ changes with varying mass and metallicity.

\subsection{Effect of the mass}
Beginning with the dependence on mass (shown in the left column of the figure), we clearly see two regimes. The lower mass models ($M \lesssim 1.55 \: {\rm{M}}_\odot$) all lie along the same line, whereas the more massive models ($M \gtrsim 1.65 \: {\rm{M}}_\odot$) show much lower values of $\mathcal{I}$ and a larger spread across different masses. Intermediate mass models transition smoothly from the higher mass (lower $\mathcal{I}$) regime to the lower mass (higher $\mathcal{I}$) regime. As the metallicity is increased, the difference between the low and high mass regimes increases. By examining the right column of Fig.~\ref{fig:mid_os_M_FeH}, we see that the $\mathcal{I}$ values in the lower mass regime depend more strongly on the metallicity.

To explain this behavior, we return to the definition of $\mathcal{I}$ given in Equation~\ref{equ:I_def}. The denominator is proportional to $(\Delta \Pi_1)^{-1}$, and so we compare the structure of different models with similar $\Delta \Pi_1$ values. We focus in Fig.~\ref{fig:N2_mass} on a small part of two evolutionary tracks with the same metallicity but different masses. The integral in the numerator of Equation~\ref{equ:I_def} is sensitive to the peak in $N^2$ caused by the chemical gradient in the hydrogen burning shell. In the models along our lower mass track in the $\Delta \Pi_1 - \mathcal{I}$ diagram, the hydrogen burning shell has already passed through any gradients in the mean molecular weight ($\mu$)  created during the main sequence evolution. Thus, the peak in $N^2$ increases due to the narrowing of the hydrogen burning shell caused by gravitational contraction, further heating the degenerate core. In contrast, the higher mass model enters our $\nu_{\rm{max}}$ regime before the burning shell has passed through the $\mu$ gradient left by the main-sequence evolution. Thus, as the star continues to evolve, the difference in $\mu$ above and below the burning shell increases rapidly, as the burning shell moves into layers that experienced less main-sequence burning. This increasing $\mu$ gradient (shown in the bottom row of Fig.~\ref{fig:N2_mass}) causes $\max N^2$ to increase rapidly until the burning shell reaches layers that were unaffected by fusion during the main sequence. From this point, the value of $\mathcal{I}$ evolves in the same manner as the lower mass track. Precisely when along each track this transition occurs depends on the maximum extent of the convective core during the main sequence. This is the first hint that using different physics when modeling will yield different values of $\mathcal{I}$ as the extent of the convective core depends on the chosen treatment of convective boundary mixing. 

It is important to note that several effects may reduce the number of stars with magnetic detections in the higher mass, lower $\mathcal{I}$ regime. Firstly, in the early part of the lower $\mathcal{I}$ sequence, the stars are still settling onto the red giant branch, with cores that are not yet fully degenerate. This means that the stars are evolving on a Kelvin-Helmholtz time scale and may therefore be difficult to detect \citep{deheuvels_seismic_2022}. However, the final part of the transition from the lower $\mathcal{I}$ sequence to the higher sequence occurs after the core is fully degenerate, and so detections in this stage may be more likely. In Appendix~\ref{app:more_figs}, we provide a plot showing the central electron degeneracy of several tracks to illustrate this point. Beyond the evolutionary time scales, magnetic signatures in stars with lower values of $\mathcal{I}$ are also inherently more difficult to detect. Equation~\ref{equ:shift_to_strength} shows that, for a given radial magnetic field strength and $\nu_{\rm{max}}$ value,  lower values of $\mathcal{I}$ result in smaller (and therefore more difficult to detect) magnetic splittings. This means that stars of higher masses require slightly stronger fields to be detectable this early in their evolution along the red giant branch. For example, a 1M$_\odot$ star with $\Delta \Pi_1 = 90$s has an $\mathcal{I}$ value roughly twice that of a 1.7M$_\odot$ star with the same metallicity and $\Delta \Pi_1$ value. Thus for the two stars to exhibit similar magnetic shifts the average radial magnetic field in the 1.7M$_\odot$ star would need to be $\sqrt{2}$ times higher than the 1M$_\odot$ star. 

Despite these two effects, we find one star in the sample of \citet{hatt_asteroseismic_2024}  (KIC~10149324) which is above the degenerate sequence in the $\Delta \Pi_1$ - $\Delta \nu$ diagram, and therefore possibly in this rapid evolutionary phase. This motivates us to use observational parameters to obtain values of $\mathcal{I}$ from our grids of models, rather than attempting to fit a relation between $\mathcal{I}$ and $\Delta \Pi_1$, as was done by \citet{li_internal_2023}. This more general approach will be even more important when magnetic signatures can be detected in higher-mass very early red giant branch stars. 

\subsection{Effect of the metallicity}

When the hydrogen burning shell has passed the $\mu$ gradient left behind by the main-sequence evolution, or for models with low enough masses to have radiative cores during the main sequence, $\mathcal{I}$ exhibits a large spread based on the metallicity of the model, see Fig.~\ref{fig:mid_os_M_FeH}. We note also that the variation of $\mathcal{I}$ with [Fe/H] decreases for increasing [Fe/H]. That is, the difference between tracks of [Fe/H] = -0.5 and 0 is larger than between tracks of [Fe/H] = 0 and +0.5, despite the fact that the difference in $Z$ is much lower in the first case (as [Fe/H] scales logarithmically with $Z$).  

\begin{figure}
    \centering
    \includegraphics{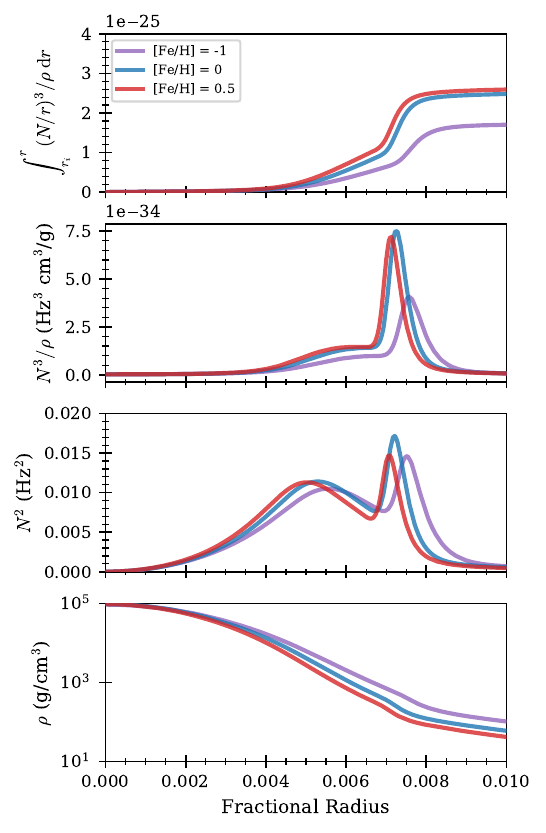}
    \caption{Dependence of $\mathcal{I}$ on metallicity for three 1M$_\odot$ models with the same $\Delta \Pi_1$. The top panel shows the cumulative integral of the numerator of Equation~\ref{equ:I_def}. The right-most value of each line shows the value of $\mathcal{I}$ for each model plotted here. The second, third, and fourth panels show the $N^3/\rho$,  $N^2$, and $\rho$ profiles, respectively.}
    \label{fig:N2_FeH}
\end{figure}

To understand this, we compare models of the same mass and $\Delta \Pi_1$ but different metallicity in Fig.~\ref{fig:N2_FeH}. Both the sub-solar ([Fe/H] = -1) and super-solar ([Fe/H] = 0.5) models have lower values of $\max N^2$ than the solar metallicity model. In the case of the sub-solar model, the lower metallicity results in less efficient CNO burning, causing a broader shell burning region and therefore a lower $\mu$ gradient. For the super-solar model, the higher metallicity reduces the change in $\mu$ across the burning shell, also reducing the $\mu$ gradient. However, because the density of the sub-solar model is higher than the super-solar model, the value of $N^3/\rho$ is significantly lower. Hence, the value of $\mathcal{I}$ is lowest for the sub-solar metallicity model. The fact that the super-solar model has a higher value of $\mathcal{I}$ may be confusing, given that the peak of $N^3/\rho$ has a higher value in the solar metallicity model. However, the integrand in the numerator of Equation~\ref{equ:I_def} is also weighted by $1/r^3$. Thus, the super-solar metallicity model’s higher values of $N^3/\rho$ in the bump below the burning shell peak lead to a slightly higher value of $\mathcal{I}$. The trend of a higher metallicity model yielding a higher value of $\mathcal{I}$ at a given $\Delta \Pi_1$ value does not hold for all masses. Specifically for higher mass tracks, the balance between the $\mu$ gradient in the burning shell and the $\rho$ profile differs, and we find that the values of $\mathcal{I}$ are lower for the [Fe/H] = 0.5 models than the [Fe/H] = 0.25 models. From this, we conclude that the metallicity is an important parameter to determine accurate values of $\mathcal{I}$.

\section{Finding best fit models to determine $\mathcal{I}$} \label{sect:best-fit}
The models examined in Sect.~\ref{sect:params} show that $\mathcal{I}$ is sensitive to several stellar properties, most notably the maximum extent of the main-sequence convective core and the metallicity. We now seek to assess what observational parameters should be considered when using best-fit models to infer $\mathcal{I}$ for a given star. In addition, we also test two different approaches to choosing the input parameters of the grid of models used. 

In order to make these assessments, we constructed a sample of 1,000 synthetic stars taken from models in our Sobol grid. The sample was chosen to have a roughly uniform distribution in mass. For each synthetic star, we calculated the likelihood weighted mean of $\mathcal{I}$ from each grid using the following sets of observed parameters: \smallskip \\ \noindent 
\begin{tabular}{l @{\hspace{4pt}} l}
    Seismic: & $\Delta \nu, \Delta \Pi_1$\\ 
    Seismic + Spectro: & $\Delta \nu, \Delta \Pi_1, T_{\rm{eff}}, {\rm{[Fe/H]}}, \sigma_{\rm{[Fe/H]}} = 0.25$ \\ 
    Seismic + HQ Spectro:& $\Delta \nu, \Delta \Pi_1, T_{\rm{eff}}, {\rm{[Fe/H]}}, \sigma_{\rm[{Fe/H]}} = 0.05$ \\ 
    Seismic + Surface:  & $\Delta \nu, \Delta \Pi_1,T_{\rm{eff}}, {\rm{[Fe/H]}}, L, \sigma_{\rm[{Fe/H]}} = 0.05$. \medskip \\ 
\end{tabular}

For the seismic parameters $\Delta \nu$ and $\Delta \Pi_1$, we adopted the mean uncertainty reported in \citet{hatt_asteroseismic_2024} of 0.01$\mu$Hz and 0.01s, respectively. We assumed a 120K uncertainty on our $T_{\rm{eff}}$ values \citep[consistent with the values reported in][]{yu_asteroseismology_2018}, and let the uncertainty of the metallicity be either 0.25 dex (the mean value reported in \citet{yu_asteroseismology_2018}) or 0.05 dex \citep[the adopted value of][]{pinsonneault_apokasc-3_2025}. For luminosity, we used an uncertainty of 2.25 $L_\odot$, which is 15\% of the mean luminosity of our grid, corresponding to the typical uncertainty reported in Gaia DR2 \citep{andrae_gaia_2018}. 
 
Although we constructed our grids with time steps roughly 1/3 of the MESA default values (by setting \texttt{time\_delta\_coeff = 0.3}), our time resolution is still not sufficiently high compared to the high precision of asteroseismic data. For example, the average change in $\Delta \Pi_1$ between successive models in our grids is over ten times the observational uncertainty. Thus, if we restrict ourselves to only the times where MESA calculated a model, we run the risk of skipping over a model that better fits the observations. While it is possible to decrease the MESA timestep such that the change in observational parameters from one model to the next is below the observational uncertainties, this is extremely computationally intensive. Instead, as is common in the literature \citep[e.g.][]{li_asteroseismology_2020, huber_stellar_2024, grusnis_tess_2025}, we opt to use interpolation to increase the effective time resolution of our grid.
We interpolate both our observed parameters and $\mathcal{I}$ values along each track onto a new time grid with steps of 10,000 years. For each synthetic star in our sample, we calculated a $\chi^2$ value for each observable parameter $P$ as 
\begin{equation}
    \label{equ:chi2}
    \chi_P^2 = \frac{(P_{\rm{obs}} - P_{\rm{model}})^2}{\sigma_P^2}, 
\end{equation}
where $P_{\rm{obs}}$ refers to the value of our synthetic star and $P_{\rm{model}}$ is the value from the models in our interpolated grid. We weighted all observables equally and thus the total $\chi^2$ value for each set of observed parameters was a sum of the individual $\chi_P^2$ values. The inferred value of $\mathcal{I}_{\rm{fit}}$ for each synthetic star was found using 
\begin{equation}
    \label{equ:infer_I} 
    \mathcal{I_{\rm{fit}}} = \frac{\sum_{i=0}^{N_{\rm{mod}}} (\mathcal{L}_i  \:  \mathcal{I}_{i, \rm{model}})}{\sum_{i=0}^{N_{\rm{mod}}} \mathcal{L}_i}, 
\end{equation}
where $\mathcal{L}_i = \exp (-\chi^2_{i, \rm{total}}/2)$ is the likelihood of the $i$th model in the interpolated grid, and the sums are taken over all $N_{\rm{mod}}$ models in each interpolated grid. When finding the value of $\mathcal{I}_{\rm{fit}}$ from the Sobol grid, we removed the track that contained the model used to generate our synthetic data. 

For each synthetic star in our sample, we calculated the error between the known value of the global magnetic sensitivity ($\mathcal{I}_{\rm{true}}$) and the one found from the grid: 
\begin{equation}
    \label{equ:delta_I} 
    \frac{\delta \mathcal{I}}{\mathcal{I}} = \frac{\mathcal{I}_{\rm{true}} - \mathcal{I}_{\rm{fit}}}{\mathcal{I}_{\rm{true}}}.
\end{equation}

\begin{figure}
    \centering
    \includegraphics[width=\linewidth]{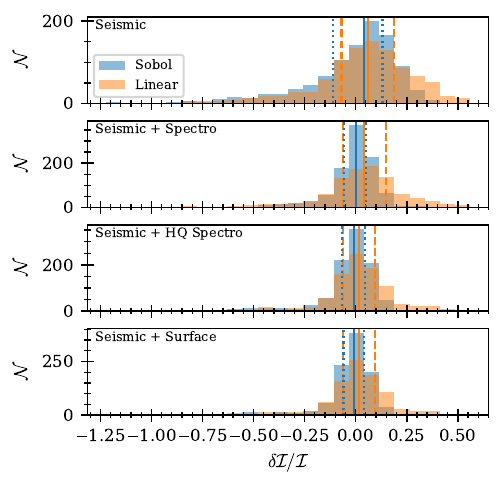}
    \caption{Comparison of the error on our inferred value of $\mathcal{I}$ when using different sets of observational parameters and grids. In each panel, we indicate the median of each grid's distribution with a solid line, and the first and third quartiles as dashed/dotted lines.}
    \label{fig:fit_grid}
\end{figure}

 For each set of observed parameters, we plot the distribution of $\delta \mathcal{I}/\mathcal{I}$ in Fig.~\ref{fig:fit_grid} from each of our two grids.  In all cases, the median on the distribution of the error in $\mathcal{I}$ is less than 10\%. However, there is a large change in the spread of our distributions (quantified by the first and third quartiles) depending on both the grid and observational parameters used. For both grids, the spread of the errors is largest when only the seismic parameters are used. Both the median error and the spread of the errors are reduced when spectroscopic variables are included in the fit, although the linear grid is more sensitive to the precision of the metallicity measurement. In all cases, global median error is similar between the two grids, however the spread is significantly smaller in our Sobol grid. We note that the typical uncertainty on $\Delta \Pi_1$ we chose based on \citet{hatt_asteroseismic_2024} is much lower than in other works. We repeated our procedure assuming an uncertainty of 0.1s on $\Delta \Pi_1$, which is the typical uncertainty of other works \citep[e.g.,][]{mosser_period_2015, vrard_period_2016, li_internal_2023, villate_seismic_2026}, and find that our results do not change. 

To understand how $\delta \mathcal{I}/\mathcal{I}$ varies with stellar parameters, we show the median, as well as the first and third quartiles, of $\delta \mathcal{I}/\mathcal{I}$ for the the Seismic + Spectro and Seismic + HQ Spectro distributions binned by the mass, [Fe/H], and $\Delta \Pi_1$ of the synthetic star in Fig.~\ref{fig:dist_summary}. The full distributions for the Seismic + HQ results can be found in Appendix~\ref{app:more_figs}. Figure~\ref{fig:dist_summary} shows that high precision metallicity values are essential to obtain $\mathcal{I}$ values for stars with low metallicity ([Fe/H] $< -0.75$), especially when using a less resolved grid, such as our Linear grid. The other important trend seen in Fig.~\ref{fig:dist_summary} is that estimates of $\mathcal{I}$ become less certain for higher mass stars, where the extent of the convective core during the main-sequence evolution becomes more important.  Based on the results shown in Figs.~\ref{fig:fit_grid}~and~\ref{fig:dist_summary}, we consider a reasonable model-based uncertainty on $\mathcal{I}$ to be 10\%, and note that extra care should be taken when examining stars with low metallicity and higher masses. We note that increasing the resolution of the model grid by running more tracks would likely reduce this uncertainty even more. However, as we explore below, in most cases the dominant source of uncertainty in the magnetic field measurement remains the measurement of the frequency splittings rather than the $\mathcal{I}$ (see Eq.~\ref{equ:shift_to_strength}), and so we do not consider the increased computational cost of running a larger grid to be necessary at this time.

\begin{figure*}
    \centering
    \includegraphics[width=\linewidth]{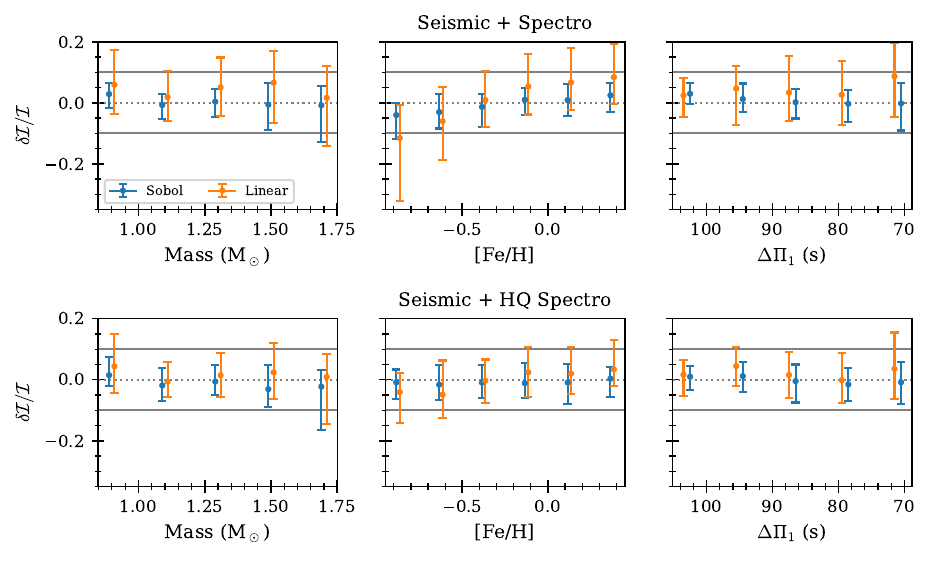}
    \caption{Summary of $\delta \mathcal{I}/\mathcal{I}$ distributions calculated using the Seismic + Spectro (top) and Seismic + HQ Spectro (bottom) sets of parameters binned by mass (left), [Fe/H] (center), and $\Delta \Pi_1$ (right). Each bin is plotted at the center of the bin, with a slight offset for visibility. The points correspond to the median of the distribution, and the error bars indicate the first and third quartiles. The gray lines show errors of $\pm 10\%$, which is our recommended value for the model-based uncertainty in $\mathcal{I}$ values determined from our Sobol grid using the Seismic + HQ Spectro set of observables.}
    \label{fig:dist_summary}
\end{figure*}

\section{Comparisons with previous work} \label{sect:compare}
\begin{figure}
    \centering
    \includegraphics{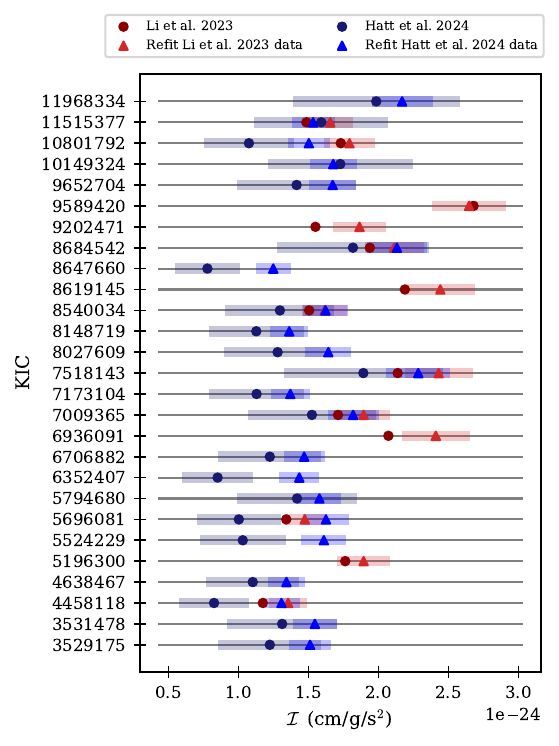}
    \caption{Comparison between the values of $\mathcal{I}$ obtained using our grid and fitting method and the $\mathcal{I}$ values in previous works. There are several stars that are studied by both \citet{li_internal_2023} and \citet{hatt_asteroseismic_2024}. As each study uses different spectroscopic measurements and obtains different seismic parameters, we fit these stars twice. Once using the stellar parameters from  \citet{li_internal_2023} (plotted as red triangles) and once with the parameters from \citet{hatt_asteroseismic_2024} (plotted as blue triangles). The shaded regions represent the uncertainties of $\mathcal{I}$ for our values and those of \citet{hatt_asteroseismic_2024}.}
    \label{fig:I_lit_comparison}
\end{figure}

We now use our Sobol grid and fitting procedure to infer values of $\mathcal{I}$ for the stars studied by \citet{li_internal_2023} and \citet{hatt_asteroseismic_2024}. Figure~\ref{fig:I_lit_comparison} shows our values of $\mathcal{I}$ and those from previous works. Our primary purpose in these comparisons is to continue exploring differences in the modeling and fitting procedure, particularly in the comparison to the observational uncertainties of the measured magnetic splittings. 

\subsection{Comparison with \citet{li_internal_2023} $\mathcal{I}$ values} 
To find our value of $\mathcal{I}$ we use the values of $T_{\rm{eff}}$, [Fe/H], $\Delta \nu$, and $\Delta \Pi_1$ reported by \citet{li_internal_2023}. We use the radial modes returned by the GYRE oscillation code \citep{townsend_gyre_2013} to calculate the $\Delta \nu$ values of our models, with more details provided in Appendix~\ref{app:MESA}. In the case of the \citet{li_internal_2023} parameters with asymmetric uncertainties, we take symmetric uncertainties based on the larger of the two values. Figure~\ref{fig:I_lit_comparison} shows that our values of $\mathcal{I}$ are consistent within our 10\% uncertainty estimates for 7 of the 13 stars examined by \citet{li_internal_2023}, and the remaining stars show differences close to 15\%. We also find that our $\mathcal{I}$ values are systematically larger than those obtained by \citet{li_internal_2023}, with only one star where our value is smaller. We attribute this bias to several factors. First, we do not use the individual mode frequencies and so are not correcting our $\Delta \nu$ values for the surface term. By using uncorrected values of $\Delta \nu$ (which are typically larger than the corrected values), the model along a given track that matches the observed $\Delta \nu$ will be more evolved than if we had applied a surface term correction, and therefore exhibit a larger value of $\mathcal{I}$. The other important difference between the grids used in this work and the grid used by \citet{li_internal_2023} is the treatment of the opacity in the stellar atmosphere.  This is important both for determining the outer boundary condition used to solve the stellar structure equations as well as the stellar atmosphere included when obtaining the oscillation frequencies. In our models, this opacity varies consistently with the local temperature and pressure. \citet{li_internal_2023} use instead a uniform opacity set by the opacity of the outermost cell in the structure model. For more details on this, we refer the reader to Sect. 6 of \citet{jermyn_modules_2023}. \citet{valle_role_2025} have previously shown that this choice significantly changes the $f_{\Delta \nu}$ correction factor applied to the large frequency scaling relations. 

To better understand the overall shift in $\mathcal{I}$ caused by these two choices, we constructed an additional set of synthetic stars from models with fixed atmosphere opacity values. Additionally, we applied the surface term prescription of \citet{li_prescription_2023} and calculated $\Delta \nu$ values from the resulting frequencies\footnote{We do not apply this prescription when fitting the stars as the relation provided in \citet{li_prescription_2023} is calibrated to a specific set of modeling choices which differ from ours.}. We then used our Sobol grid to find the best-fit value of $\mathcal{I}$ for this new set of synthetic stars. In contrast to the results presented in Sect.~\ref{sect:best-fit}, our grid overestimates $\mathcal{I}$ by around 5\% in the mass and $\Delta \Pi_1$ regimes covered by the \citet{li_internal_2023} sample. From this, we conclude that the differences between our values of $\mathcal{I}$ and those of \citet{li_internal_2023} likely stem from our different treatment of the opacity in the stellar atmosphere and our use of $\Delta \nu$ values that have not been corrected by the surface term. For future work, we recommend applying surface term corrections carefully and using models with a more realistic treatment of atmospheric opacity. 

\subsection{Comparison with \citet{hatt_asteroseismic_2024}}
We focus here on the stars with significant detections of magnetic splittings ($\delta \nu_{\rm{mag}} > 2\sigma_{\delta \nu_{\rm{mag}}}$) reported in \citet{hatt_asteroseismic_2024}. We use their reported values of $T_{\rm{eff}}$, [Fe/H], and $\Delta \Pi_1$. Rather than using their reported values of $\Delta \nu$, we use the radial mode frequencies directly. This allows us to correct for the surface term using the two-term formulation of \citet{ball_new_2014}, where the free parameters are determined independently for each model in our interpolated grid. We treat the frequencies as a single observable, that is 
\begin{equation}
    \chi^2_{\rm{freq}} = \frac{1}{N_{\rm{freq}}} \sum_i^{N_{\rm{freq}}} \frac{(\nu_{i, \rm{obs}} - \nu_{i, \rm{mod}})^2}{\sigma_{\nu_i}^2},
\end{equation} 
where $N_{\rm{freq}}$ is the number of frequencies and $\nu_{i,\rm{mod}}$ is the model frequency after correcting for surface effects. Although not fully correct in a statistical sense, weighting a frequency $\chi^2$ by the number of frequencies is common in the asteroseismic literature to avoid the individual frequencies overwhelming any contribution from other global parameters. For a longer discussion of this topic, see \citet{cunha_plato_2021}. 

When comparing our values of $\mathcal{I}$ to those obtained by \citet{hatt_asteroseismic_2024}, we again find that our values are systematically larger\footnote{In addition, we find one star (KIC~9508757) where our method did not return a value of $\mathcal{I}$. This is because this star is below the $\Delta \nu - \Delta \Pi_1$ sequence expected from single star evolution. This suggests that KIC~9508757 may be the product of binary interactions \citep{deheuvels_seismic_2022}. As we do not account for these effects, we do not include this star in our comparisons.}. In addition the difference between our values are larger with an average difference of 20\%. Although this is below the 30\% uncertainty on $\mathcal{I}$ adopted by \citet{hatt_asteroseismic_2024}, it is larger than our adopted 10\% uncertainties. Additionally, we find six stars where our values of $\mathcal{I}$ differ from those in \citet{hatt_asteroseismic_2024} by more than 40\%. 

The reasons for these large differences are less clear than the differences discussed in the previous section. Again, our choice to vary the opacity in the atmosphere likely plays a role in pushing our results to higher values of $\mathcal{I}$. Also potentially playing a role is the difference in the grid construction. Since we use our Sobol grid, $\alpha_{\rm{mlt}}$, Y$_{\rm{init}}$, and $f_{ov}$ are free parameters. In the grid used by \citet{hatt_asteroseismic_2024} $\alpha_{\rm{mlt}}$ and $f_{ov}$ are fixed and Y$_i$ is varied linearly with Z$_{\rm{init}}$. We see large differences in $\mathcal{I}$ for stars with masses low enough to have radiative cores on the main sequence, and so the bias we see cannot be fully explained by our variation of the $f_{ov}$. The treatments of  $\alpha_{\rm{mlt}}$ and Y$_{\rm{init}}$ are known to change the stellar parameters inferred using seismic modeling \citep[e.g.][]{li_realistic_2024}, and it is perhaps unsurprising that our different grid construction techniques result in different values of $\mathcal{I}$. However, we were unable to find a satisfying explanation for why our choice to vary these parameters should yield systematically higher values of $\mathcal{I}$. However, for the stars analyzed by both \citet{li_internal_2023} and \citet{hatt_asteroseismic_2024}, the values of $\mathcal{I}$ reported by \citet{li_internal_2023} are larger in all but one case. As the grid used by \citet{li_internal_2023} also varies $\alpha_{\rm{mlt}}$ and Y$_{\rm{init}}$, this supports the idea that the grid construction technique can lead to systematically larger values of $\mathcal{I}$.

The final difference between our method and the methods of both \citet{li_internal_2023} and \citet{hatt_asteroseismic_2024} that is important to note, although difficult to predict the effect of, is that we interpolate along the evolutionary tracks in our grid. This means that we are able to increase the effective time resolution of our grid without the additional computational expense of running MESA with smaller time steps. In theory, one could also interpolate between tracks (increasing the resolution in the varied initial stellar and physical parameters), although we did not pursue this. 

\subsection{Comparison of radial magnetic field strength values and their uncertainties} 
\begin{figure}
    \centering
    \includegraphics{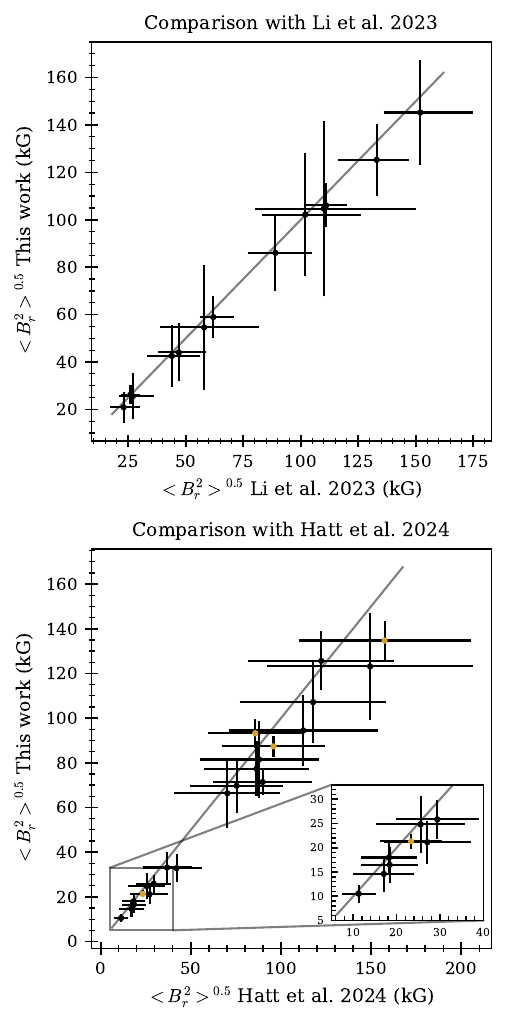}
    \caption{Comparison of our inferred average magnetic field strengths with those of previous studies. In both panels, we plot the 1:1 line in gray. In the lower panel, the yellow points indicate stars where our model-based uncertainty of $\mathcal{I}$ is larger than the uncertainty on the observed magnetic splittings. }
    \label{fig:B_lit_comparison}
\end{figure}

Using the magnetic frequency shifts found by \citet{li_internal_2023} and \citet{hatt_asteroseismic_2024}, we calculated the resulting average magnetic field strengths using our values of $\mathcal{I}$ according to Equation~\ref{equ:shift_to_strength}. Figure~\ref{fig:B_lit_comparison} compares our results to those of the previous works. Due to the higher values of $\mathcal{I}$ that we find, our magnetic field strengths are generally lower than in previous works. As we did not correct for surface effects when fitting the stars in \citet{li_internal_2023}, our values should not be taken as corrections to the previous values, and so we do not provide them. In the case of the sample of \citet{hatt_asteroseismic_2024}, we were able to correct for surface effects, and so we provide our new values in Table~\ref{tab:Hatt_new_B_values}. We find that although our values of $\mathcal{I}$ differ, the final field strength values agree within the final uncertainties for all of the stars in \citet{li_internal_2023} and  \citet{hatt_asteroseismic_2024}. This is unsurprising as, for most of the stars, the uncertainty in the final result is dominated by the uncertainty of the measured magnetic shift. In the \citet{hatt_asteroseismic_2024} sample, there are four stars where the percent uncertainty of the magnetic splittings is below 10\%. For these four stars, the model-based uncertainty of $\mathcal{I}$ is larger than the uncertainty on the average magnetic field strength. This shows that although most measurements are not yet precise enough for the stellar modeling to be the dominant source of uncertainty, in stars with the most precisely measured magnetic shifts, the modeling choices used to obtain $\mathcal{I}$ can be significant.

\begin{table} 
    \caption{Global magnetic sensitivities and average radial magnetic field strengths obtained from our procedure for stars in \citet{hatt_asteroseismic_2024}. }
    \label{tab:Hatt_new_B_values}
    \centering 
    \begin{tabular}{c c c}
    \hline \hline 
     KIC & $\mathcal{I}\:$ ($10^{-24}$ cm/g/s$^{2}$) & $\left<B_r^2\right>^{0.5}$(kG) \\  \hline
    3529175 & 1.489 $\pm$ 0.149 & 107.2$\pm$18.3 \\
    3531478 & 1.532 $\pm$ 0.153 & 69.7$\pm$12.4 \\
    4458118 & 1.292 $\pm$ 0.129 & 94.4$\pm$15.7 \\
    4638467 & 1.338 $\pm$ 0.134 & 81.5$\pm$17.4 \\
    5524229 & 1.721 $\pm$ 0.172 & 21.2$\pm$4.4 \\
    5696081 & 1.359 $\pm$ 0.136 & 134.7$\pm$8.8 \\
    5794680 & 1.614 $\pm$ 0.161 & 24.8$\pm$5.9 \\
    6352407 & 1.358 $\pm$ 0.136 & 32.8$\pm$6.2 \\
    6706882 & 1.251 $\pm$ 0.125 & 125.6$\pm$13.2 \\
    7009365 & 1.836 $\pm$ 0.184 & 77.3$\pm$12.4 \\
    7173104 & 1.488 $\pm$ 0.149 & 33.1$\pm$7.0 \\
    7518143 & 2.28 $\pm$ 0.228 & 21.3$\pm$1.5 \\
    8027609 & 1.622 $\pm$ 0.162 & 66.3$\pm$15.5 \\
    8148719 & 1.649 $\pm$ 0.165 & 14.6$\pm$3.6 \\
    8540034 & 1.627 $\pm$ 0.163 & 25.8$\pm$3.9 \\
    8647660 & 1.246 $\pm$ 0.125 & 123.1$\pm$23.9 \\
    8684542 & 2.187 $\pm$ 0.219 & 87.3$\pm$4.6 \\
    9652704 & 1.794 $\pm$ 0.179 & 16.5$\pm$3.6 \\
    10149324 & 1.791 $\pm$ 0.179 & 18.0$\pm$3.5 \\
    10801792 & 1.696 $\pm$ 0.17 & 71.4$\pm$5.8 \\
    11515377 & 1.355 $\pm$ 0.136 & 93.2$\pm$6.5 \\
    11968334 & 2.289 $\pm$ 0.229 & 10.5$\pm$1.8 \\
    \hline
\end{tabular}
\end{table}

\section{Conclusions} \label{sect:conc} 
In this work, we used stellar models to explore the global magnetic sensitivity, $\mathcal{I}$, used to infer internal magnetic field strengths in red giants. We looked at how $\mathcal{I}$ varies with stellar parameters, as well as how well $\mathcal{I}$ can be recovered using grid-based techniques and different observational constraints. We found that the mass dependence of $\mathcal{I}$ is largest for models where the hydrogen burning shell is still below the mean molecular weight gradient caused by the recession of the convective core during main sequence evolution. After this point, or in models with radiative cores on the main sequence, the value of $\mathcal{I}$ depends more on the metallicity than the mass.

To test how accurately $\mathcal{I}$ can be determined from observed parameters, we constructed a set of synthetic stars and attempted to infer $\mathcal{I}$ from our model grids. We tested several different sets of observed parameters and found that for all sets the median error on $\mathcal{I}$ is less than 5\%, with spreads between 10\% and 15\%. We obtained the best results when our fitting procedure included high-quality spectroscopic data, particularly [Fe/H] values, in addition to the global seismic parameters $\Delta \nu$ and $\Delta \Pi_1$. Based on our tests, we recommended adopting a 10\% uncertainty on $\mathcal{I}$ when found using this method. 

In comparison to previously reported values of $\mathcal{I}$, our procedure yields systematically higher values of $\mathcal{I}$. In the case of the stars studied by \citet{li_internal_2023}, we attribute this primarily to a difference in the treatment of atmospheric opacity in our models and our use of model $\Delta \nu$ values without applying a surface term correction. We attribute the large differences in $\mathcal{I}$ between this work and \citet{hatt_asteroseismic_2024} to our differing grid construction approaches, notably the treatment of the mixing length and overshoot parameters as well as the initial helium abundance. We note that while we have focused on red giant stars, a similar approach has been proposed to measure internal magnetic fields in main-sequence f-type stars \citep{takata_asteroseismic_2025}. Our results suggest that it may be important to account for model-based uncertainties in these stars as well. 

One important limitation of this work is that we only examined single-star evolution. Some stars with measured magnetic splittings show evidence of binary interactions \citep{deheuvels_seismic_2022, villate_seismic_2026}. These interactions are known to change the internal structure of the star \citep{rui_asteroseismic_2021} and hence the value of $\mathcal{I}$ (see Muntean et al., in prep). For stars where single star evolution is appropriate, however, our grid and fitting method provide reliable estimates of $\mathcal{I}$ with uncertainties below the uncertainty of typical observations. To aid future measurements of internal magnetic fields in red giants, we provide both our Sobol grid and our scripts to infer values of $\mathcal{I}$ at \url{https://zenodo.org/records/21032697}. 

%%%%%%%%%%%%%%%%%%%%%%%%%%%%%%%%%%%%%%%%%%%%%%%%%%%%%%%%%%%%%%
\begin{acknowledgements}
L. Buchele and L. Bugnet gratefully acknowledge support from the European Research Council (ERC) under the Horizon Europe programme (Calcifer; PI Bugnet; Starting Grant agreement N$^\circ$101165631). While partially funded by the European Union, views and opinions expressed are, however, those of the authors only and do not necessarily reflect those of the European Union or the European Research Council. Neither the European Union nor the granting authority can be held responsible for them.
\end{acknowledgements}

%%%%%%%%%%%%%%%%%%%%%%%%%%%%%%%%%%%%%%%%%%%%%%%%%%%%%%%%%%%%%%
\bibliographystyle{aa} % style aa.bst
\bibliography{references} % your references Yourfile.bib
%%%%%%%%%%%%%%%%%%%%%%%%%%%%%%%%%%%%%%%%%%%%%%%%%%%%%%%%%%%%%%

%%%%%%%%%%%%%%%%%%%%%%%%%%%%%%%%%%%%%%%%%%%%%%%%%%%%%%%%%%%%%%%
% Appendices must be placed after   \end{thebibliography}
% They will be placed automatically on a new page.
%%%%%%%%%%%%%%%%%%%%%%%%%%%%%%%%%%%%%%%%%%%%%%%%%%%%%%%%%%%%%%%
\begin{appendix} 
\nolinenumbers
\section{Details of MESA calculations} \label{app:MESA} 
Here we provide more details about our MESA setup, including the model physics we kept constant between our two grids and our calculation of the asteroseismic parameters. We provide the MESA and GYRE inlists we used, as well as our \texttt{run\_star\_extras} files at the Zendo link provided above. 

\subsection{Model physics} 
As noted in the main text, most of our physics choices follow those in \citet{li__beyond_2025}. We use the default blend of equation of state data from OPAL \citep{rogers_updated_2002}, FreeEOS \citep{irwin_freeeos_2004}, and Skye \citep{jermyn_skye_2021}. We use the solar abundance scale of \citet{grevesse_standard_1998} to scale the initial metal abundances of our models and also when calculating their [Fe/H] values. Our radiative opacity values are taken from OPAL \citep{iglesias_radiative_1993, iglesias_updated_1996} and supplemented in the low-temperature regime by values from \cite{ferguson_low-temperature_2005}. We adopt the cubic interpolation scheme for our opacity values described in \citet{farag_expanded_2024}. Our nuclear network choice is \texttt{pp\_cno\_extras\_o18\_ne22.net} with reaction rates taken from JINA REACLIB \citep{cyburt_jina_2010}, NACRE \citep{angulo_compilation_1999}, as well as additional tabulated weak reaction rates \citep{fuller_stellar_1985, oda_rate_1994, langanke_shell-model_2000}. 

We use the mixing length formulation of \citet{henyey_studies_1965}. We determine convective boundaries using the Ledoux criterion and turn on MESA's convective premixing scheme \citep{paxton_modules_2019}. Although the adopted value of the overshooting parameter varies between our models, all models use exponential overshooting \citep{herwig_evolution_2000} with $f_0 = 0.004$. We follow the recommendation of \citet{buchele_exploring_2025} and set \texttt{overshoot\_D\_min} to $10^{-2}$ cm$^2$/s. In addition, we implement a routine to ensure that extra mesh points are placed around the overshooting regions. For our atmospheric boundary conditions, we adopt the Eddington T-$\tau$ relation with a varying opacity consistent with the local temperature and pressure, except during pre-main-sequence evolution, where adopting a fixed opacity improved the stability of the models.

\subsection{Calculation of asteroseismic parameters} 
Since there are several ways of obtaining seismic parameters from stellar models, we provide the details of our calculations here. 
\subsubsection{$\Delta \nu$} 
To obtain a value of $\Delta \nu$ for each model in our grid, we calculate the radial mode frequencies using GYRE \citep{townsend_gyre_2013}. We obtain frequencies in the range $0.001\mu$Hz to $2\nu_{ac}$ where $\nu_{ac}$ is the acoustic cutoff frequency of the model. Defining the frequency range in this manner allows us to call GYRE on-the-fly as MESA is running rather than as a post-processing step. However, this range of frequencies is far larger than what is observed in these stars. To account for this, we calculate $\Delta \nu$ using a weighted least squares fit between the frequencies and their radial orders. Our weighting function is a Gaussian centered at $\nu_{\rm{max}}$ with a FWHM of $0.66\nu_{\rm{max}}^{0.88}$, which is the scaling of red giant oscillation power excess found by \citet{mosser_characterization_2012}.

\subsubsection{$\Delta \Pi_1$ and $\mathcal{I}$}
\begin{figure}
    \centering
    \includegraphics{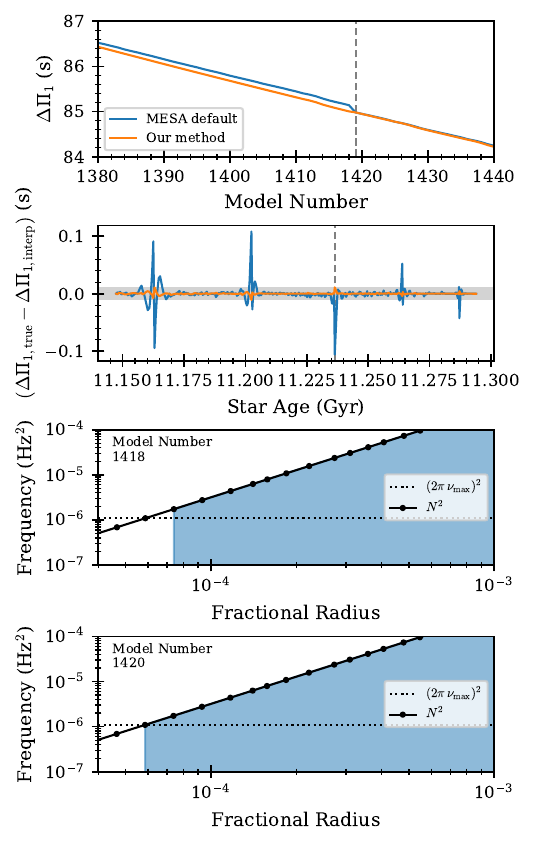}
    \caption{Comparison of the MESA default way of calculating $\Delta \Pi_1$ and our method. The top panel shows the evolution of $\Delta \Pi_1$ calculated with each method. The second panel shows the interpolation error from each method, with the gray shaded region indicating the average uncertainty of $\Delta \Pi_1$ reported in \citet{hatt_asteroseismic_2024}. The bottom two panels show, in the blue areas, the region used by MESA to calculate $\Delta \Pi_1$ for two models on either side of the jump indicated by the dashed vertical line in the top two panels.}
    \label{fig:DPi_mesh_issues}
\end{figure}

Although we use GYRE to calculate the radial mode frequencies, calculating mixed modes is significantly more computationally expensive, and so we opt to use the asymptotic form of $\Delta \Pi_1$ given in Equation~\ref{equ:DPi}. While this is a quantity that MESA can provide, we find that the values of $\Delta \Pi_1$ outputted by the default routine do not evolve smoothly along a given track. Instead, there are points where the value jumps abruptly from one model to the next, see the top panel of Fig.~\ref{fig:DPi_mesh_issues}. This occurs because the integration limits are defined by $N^2$ crossing $(2 \pi \, \nu_{\rm{max}})^2$; however, there is no guarantee that MESA places a mesh point at the exact point where this crossing occurs. Thus, $\Delta \Pi_1$ jumps when the integration limits of Equation~\ref{equ:DPi} are expanded by one zone. Although we show this for only for the inner turning point in the lower two panels of Fig.~\ref{fig:DPi_mesh_issues}, we find similar issues (of smaller magnitude) at the outer turning point. These numerical issues cause $\Delta \Pi_1$ to be slightly overestimated for a period leading up to the jump and also introduce significant interpolation error, shown in the second panel of Fig.~\ref{fig:DPi_mesh_issues}. To resolve this issue, we interpolate $N^2$ around both the inner and outer turning points to find the exact radius where $N^2 = (2 \pi \,\nu_{\rm{max}})^2$. We then include these partial zones in our calculation of $\Delta \Pi_1$. This procedure is also applied to the integrals used to calculate $\mathcal{I}$, see Equation~\ref{equ:I_def}.

\section{Additional Figures} \label{app:more_figs}
\begin{figure} 
    \centering
    \includegraphics[width=\linewidth]{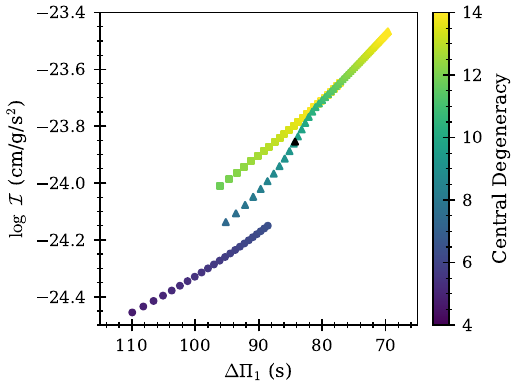}
    \caption{Change in central electron degeneracy (described as the Fermi energy in units of $k_b T$) for three different solar metallicity tracks from our Linear grid. The marker indicates the mass: 1M$_\odot$ (square), 1.55M$_\odot$ (triangle), 1.6M$_\odot$ (circle). The black triangle indicates the point where the 1.55M$_\odot$ track joins the degenerate sequence in the $\Delta \nu$ -- $\Delta \Pi_1$ diagram. }
    \label{fig:center_deg}
\end{figure}

\begin{figure}
    \centering
    \includegraphics[width=\linewidth]{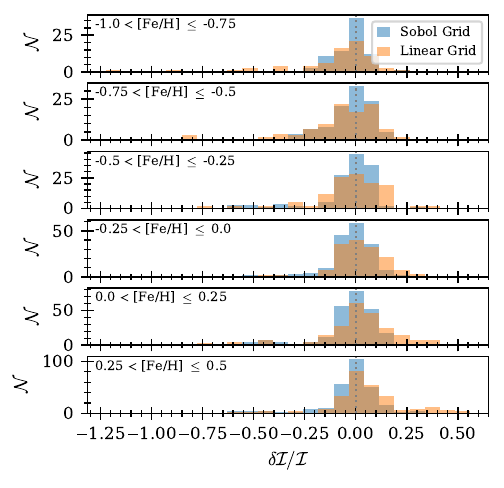}
    \caption{Comparison of the results best fit models found using Seismic + HQ Spectra variables for different values of the synthetic star's [Fe/H].}
    \label{fig:fit_FeH}
\end{figure}

\begin{figure}
    \centering
    \includegraphics[width=\linewidth]{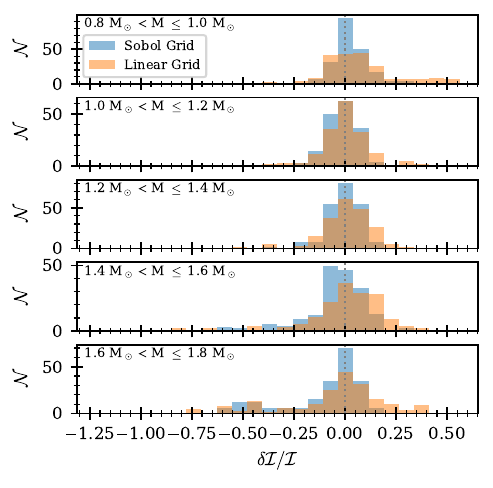}
    \caption{Comparison of the results best fit models found using Seismic + HQ Spectra variables for different values of the synthetic star's mass.}
    \label{fig:fit_mass}
\end{figure}

\begin{figure}
    \centering
    \includegraphics[width=\linewidth]{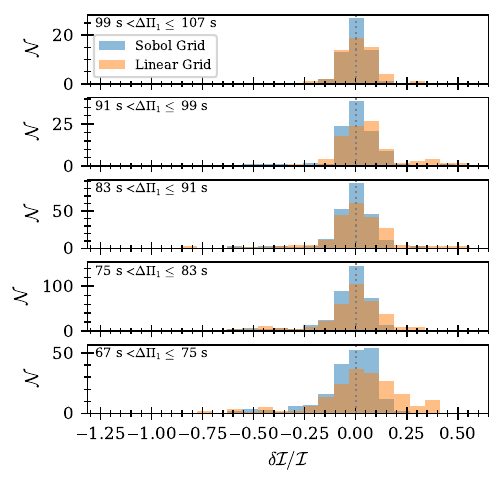}
    \caption{Comparison of the results best fit models found using Seismic + HQ Spectra variables for different values of the synthetic star's $\Delta \Pi_1$.}
    \label{fig:fit_DPi}
\end{figure}

Here we provide a few additional figures for the interested reader. In Fig.~\ref{fig:center_deg}, we show how the central electron degeneracy changes for several solar metallicity tracks of various masses. In Figs.~\ref{fig:fit_mass}, \ref{fig:fit_FeH}, and \ref{fig:fit_DPi}, we provide the full distributions of $\delta \mathcal{I}/\mathcal{I}$ found when fitting using the Seismo + HQ Spectro observables binned by mass, [Fe/H], and $\Delta \Pi_1$ that were summarized in Fig.~\ref{fig:dist_summary}.

\end{appendix}
\end{document}